\documentclass[twocolumn,twoside,slac_two]{revtex4}
\usepackage{graphicx}
\usepackage{fancyhdr}
\usepackage{url}
\usepackage{amsmath}
\usepackage{ulem}
\usepackage{natbib}
\newcommand{\degr}{^{\circ}}

\begin{document}

\title{The {\it Fermi}-LAT model of interstellar emission for standard point source analysis}
\author{Jean-Marc Casandjian for the {\it Fermi} Large Area Telescope Collaboration}
\affiliation{Laboratoire AIM, CEA-IRFU/CNRS/Universit\'e Paris Diderot, Service d'Astrophysique, CEA Saclay, 91191 Gif sur Yvette, France}
\email{casandjian@cea.fr}
\date{\today\\Version 2}

\begin{abstract}
We describe the development of the model for interstellar $\gamma$-ray emission that is the standard adopted by the LAT team and is publicly available. The model is based on a linear combination of templates for interstellar gas column density and for the inverse Compton emission. The spectral energy distributions of the $\gamma$-ray emission associated with each template are determined from a fit to 4 years of {\it Fermi}-LAT data in 14 independent energy bins from 50 MeV to 50 GeV. We fit those distributions with a realistic model for the emission processes to extrapolate to higher energies. We also include large-scale structures like Loop~{\sc i} and the {\it Fermi} bubbles following an iterative procedure that re-injects filtered LAT counts residual maps into the model. We confirm that the cosmic-ray proton density varies with the distance from the Galactic center and find a continuous softening of the proton spectrum with this distance. We observe that the {\it Fermi} bubbles have a shape similar to a catenary at their bases.
\end{abstract}

\maketitle
\thispagestyle{fancy}

\section{Principle}
\label{sec:Introduction_section}
This paper describes the model of interstellar emission recommended for point source analyses of the {\it Fermi}-LAT Pass 7 reprocessed data (P7REP) \citep{Bregeon:2013p4093}. The high-energy interstellar $\gamma$-ray emission is produced by the interaction of energetic cosmic rays (CRs) with interstellar nucleons and photons. The decay of secondary particles produced in hadron collisions, the inverse Compton scattering of the interstellar radiation field (ISRF) by electrons and their bremsstrahlung radiation emission in the interstellar medium (ISM) are the main contributors to the Galactic diffuse emission. The interstellar emission model is based on a template method: we assumed that the diffuse $\gamma$-ray intensity at any energy can be modeled as a linear combination of maps or templates of hydrogen column-density ($I_{H}$) and a predicted inverse Compton (IC$_p$) intensity map ($I_{IC_{p}}$) calculated by GALPROP \footnote{http://galprop.stanford.edu}. The intensity of each template in each energy bin is determined from a fit to the P7REP Clean class events from the first 4 years of the mission binned into 14 equal logarithmic intervals from 50~MeV to 50~GeV. \\
In addition to the interstellar emission, the LAT detects $\gamma$ rays from other sources that need to be taken into account in the analysis. We do this by adding dedicated components to account for a residual intensity of the Earth limb ($I_{limb}$), for point and extended $\gamma$-ray sources ($I_{ext}$), and for the emission from the Sun and the Moon (${I}_{sun\_moon}$). Finally we add an uniform intensity template ($I_{iso}$) to account for unresolved $\gamma$-ray sources and CR contamination in the data. For a given counts map pixel and energy band we calculated the predicted number of counts ($N_{pred}$) detected by {\it Fermi}-LAT as:

{\footnotesize
\begin{equation}
\begin{split}
N_{pred}(E) = \sum_{i=H~templates} q_{i}(E)\widetilde{I}_{H_i}  +  N_{IC}(E)\widetilde{I}_{IC_{p}}(E) \\  +  N_{iso}(E)\widetilde{I}_{iso}  + N_{LoopI}(E)\widetilde{I}_{LoopI} + \sum_{i=patch} N_{patch_i}(E)\widetilde{I}_{patch_i}    \\    + N_{limb}(E)\widetilde{I}_{limb}   + \sum_{i=point~src} N_{pt_{i}}(E)\widetilde{\delta}(i)  \\ +  \sum_{i=extend~src} N_{ext_{i}}(E)\widetilde{I}_{ext_i} +  \widetilde{I}_{sun\_moon}(E) 
\end{split}
\label{eqRing}
\end{equation}
}
where E is the energy. We use the notation $\widetilde{I}$ to denote predicted counts maps resulting from the convolution by the LAT PSF of the product of an intensity map $I$ and the instrument exposure and pixel solid angle. In Equation \ref{eqRing} each template of hydrogen column density $I_{H}$ is multiplied by its associated hydrogen $\gamma$-ray emissivity $q$. The factors $N_{IC}$, $N_{limb}$, and $N_{iso}$ represent the renormalization factors associated to $I_{IC_{p}}$, $I_{limb}$, and $I_{iso}$. ${I}_{sun\_moon}$ is kept fixed in the analysis. Equation \ref{eqRing} also incorporates coefficients associated to extended sources ($N_{ext}$) and to point sources ($N_{pt}$) represented by the Dirac $\delta$ function. $I_{LoopI}$ and  $N_{LoopI}$ account for local IC emission from Loop~{\sc i}. For unmodeled excesses, we also introduced in Equation \ref{eqRing} patches of uniform intensity ($I_{patch}$) with normalization factors $N_{patch}$. The procedure behind the construction of our interstellar emission model was to find templates for the gas column density and IC$_{p}$, fit Equation \ref{eqRing} to {\it Fermi}-LAT counts maps with  $q$, $N_{IC}$, $N_{iso}$, $N_{LoopI}$, $N_{patch}$, $N_{limb}$, $N_{pt}$ and $N_{ext}$ left free to vary in each energy bin and to extrapolate the coefficients related to the hydrogen templates and $IC_{p}$ outside the energy range of the fit. \\

\section{Template description }
About 99\% of the ISM mass is gas and about 70\% of this mass is hydrogen. The hydrogen gas exists in the form of neutral atoms in cold and warm phases, in the form of neutral molecules (H$_2$), and in an ionized state. Helium and heavier-elements are assumed to be uniformly mixed with the hydrogen. \\
H~{\sc i} is traced by its radio 21-cm line radiation; we derived its column density $N$H~{\sc i} from the 21-cm line radiation temperatures under the assumption of a uniform spin (excitation) temperature ($T_{S}$). The 21-cm all-sky Leiden-Argentine-Bonn (LAB) composite survey of Galactic H~{\sc i} \citep{Kalberla:2005p3048} is used to determine the all-sky distribution of $N$H~{\sc i}. We derived $N$H~{\sc i} from the observed brightness temperature using a $T_{S}$ of 140~K which provided the best fit to the {\it Fermi}-LAT data in regions with $90\degr \le l \le 270\degr$ and $\left| b \right| <70\degr$. Because the CR flux varies with Galactocentric distance and Equation \ref{eqRing} is valid only if the CR flux is uniform in each template, we partitioned the Galaxy into Galactocentric annuli and assign to each annulus the corresponding H~{\sc i} column density. \\
The molecular hydrogen which does not have a permanent dipole moment, generally can not be observed directly in its dominantly cold phase. The observation of molecular gas relies on other molecules and especially on the 2.6-mm J=1$\rightarrow$0 line of carbon 12 monoxide (CO). It is common to assume that the H$_2$ column density is proportional to the velocity-integrated CO brightness temperature $W$(CO). The molecular hydrogen-to-CO conversion factor is expressed as $X_{CO}$=$N$(H$_{2}$)/$W$(CO). We obtained the $W$(CO) spatial distribution from the Center for Astrophysics composite survey \citep{Dame:2001p1849}. We derived Galactocentric annuli from radial CO velocities in a similar way as for H~{\sc i}. \\
Unfortunately CO is not a perfect tracer of H$_2$. Moreover $N$H~{\sc i} derived under the hypothesis of a uniform $T_{S}$=140~K is likely to be biased in lines of sight crossing regions of different $T_{S}$. Those approximations lead to large underestimates of the quantities of gas called dark neutral medium (DNM) in our Galaxy \citep{Grenier:2005p836, Ade:2011p3178, Paradis:2012p4067}. Since dust is well mixed with gas, we accounted for this gas by including in our model a template related to the total dust column density. We derived a DNM template as the residual map obtained after subtracting from the dust optical depth map of the \cite{Schlegel:1998p290} parts linearly correlated with the $N$H~{\sc i} and $W$(CO) annuli. Subtracting the correlated parts from the dust optical depth map revealed coherent structures across the sky both in the positive and in the negative residuals. The negative residuals are likely related to regions in which an average $T_{S}$ of 140~K is too low, and thus $N$H~{\sc i} is overestimated. In this paper we call this map the ``$N$H~{\sc i} correction map''. The positive residuals reveal gas in addition to that traced by $N$H~{\sc i} and $W$(CO). Here we associated this excess map to the DNM distribution even if it also includes regions in which an average $T_{S}$ of 140~K is too high and potentially incorporates also ionized hydrogen that might be mixed with dust. \\
Due to their proximity, CR protons and electrons interacting with the Earth limb make the Earth by far the brightest $\gamma$-ray source in the sky \citep{Abdo:2009p4150}.  The {\it Fermi} standard observational strategy is such that the Earth is not directly in the field of view of the LAT.  However a large number of limb photons entering the LAT at large zenith angles are still detected. We constructed a simple template based on the subtraction of the counts map derived with a zenith angle cut at 100$\degr$ to a counts map restricted to angles above 80$\degr$. \\ 
In fitting the model for interstellar diffuse emission we included templates for 21 extended sources and 2179 point sources at positions listed in a first iteration of the third {\it Fermi}-LAT catalog (3FGL) \citep{Ballet:2013p4107} derived with a preliminary iteration of the Galactic diffuse emission model. We also incorporated in the fit the $\gamma$-ray emission from the Sun and the Moon. Their intensities were not allowed to vary during the $\gamma$-ray template fit procedure. \\
While the different gas column-density maps offer templates for photons originating mainly from $\pi^{0}$-decay and Bremsstrahlung emission, there is no direct observational template for the IC emission. Instead it must be calculated. For that we used the prediction from the GALPROP code with GALDEF identification $^SY^Z6^R30^T150^C2$, a representative diffusive reacceleration model described in \cite{Ackermann:2012p2978}. \\
Excesses originating for example from Loop~{\sc i} \citep{Casandjian:2009p3311,Ackermann:2012p2978} or the {\it Fermi} Bubbles \citep{Su:2010p2675,Ackermann:2014p4189} are observed when we compared a preliminary template model derived only from the templates mentioned above to the {\it Fermi}-LAT observations. There is no accurate a priori template for the $\gamma$-ray emission of those large structures. Not including them in Equation \ref{eqRing}, as well as other structures that we observed in the residuals at lower latitudes, will strongly bias the fit. To reduce this bias we roughly modeled the strongest emitting region of Loop~{\sc i} with a selected region of the 408~MHz radio continuum intensity from the survey of \cite{Haslam:1982p4092}. To account for excesses not correlated with radio templates we introduced ad hoc patches in Equation \ref{eqRing}. The patches are regions of spatially uniform intensity whose shapes encompass regions with an excess of photons of at least about 20\% compared to the preliminary model. We added 4 patches including a large rounded shape filling Loop~{\sc i}  and three smaller patches closer to the Galactic plane. Additionally, we have created two patches for the {\it Fermi} bubbles. We also made a disk-shaped patch around the Cygnus region. We used the patches to derive the $\gamma$-ray emissivities of the hydrogen that we deduced from the fit of Equation \ref{eqRing} to the {\it Fermi}-LAT data. But the patches do not provide an accurate enough description of the interstellar emission to be added to the final interstellar model. Instead we have incorporated a filtered $\gamma$-ray residuals map obtained with the patches removed as described below.

\section{Gamma-ray fit and emissivity interpretation}
We did not fit Equation \ref{eqRing} to the {\it Fermi}-LAT all-sky at once, we applied latitude and longitude cuts to define subsets corresponding to regions where some templates are prominent.
We fitted the local atomic hydrogen in the whole longitude range but away from the Galactic plane at latitudes $|b|>10\degr$. We lowered this latitude cut for local CO annulus to 4$\degr$ and to 3$\degr$ for the $N$H~{\sc i} correction maps and  DNM template obtained from the negative and positive dust residual. For the annuli with Galactocentric radii larger than 10~kpc (H~{\sc i} and CO outer Galaxy annuli) we applied only a longitude cut corresponding to $90\degr<l<270\degr$. For those 4 independent fits we left all the template normalization coefficients of Equation \ref{eqRing} free to vary in each of the 14 energy bins except for the Sun and the Moon templates. \\
The inner Galactic region is particularly difficult to model. The gas column densities are affected by optical depth correction, self-absorption of H~{\sc i} and limited kinematic distance resolution at low longitudes. Additionally $\gamma$-ray point and extended sources are numerous, a DNM template is lacking, and the IC$_{p}$ morphology is uncertain. Possibly for one of those reasons, or because of an excess of CR or an incorrect modeling of a foreground emission, we observed a broad unmodeled emission (referred to as ``extra emission'' in the rest of the text) in the direction of the inner Galaxy with a maximum in the first Galactic quadrant at a longitude of $\sim$30$\degr$. Up to this stage this emission was approximately accounted for by patches of uniform intensity. At this point we removed the patches in Equation \ref{eqRing} and we modeled the extra emission with a two-step iterative procedure: \\
In the first step we fitted Equation \ref{eqRing}  to the {\it Fermi}-LAT observations excluding  $5\degr<l<90\degr$ and  $\left| b \right| <20\degr$. We obtained a residual map with some emission not accounted for, we selected the positive residuals, we smoothed them with a 2-dimensional Gaussian symmetric  kernel of 3$\degr$ FWHM and re-injected them in Equation \ref{eqRing} as a template for another iteration. We iterated three times up to the point where the positive residual intensities approximately equalled the intensities of negative ones. We obtained a first set of $\gamma$-ray emissivities in the inner H~{\sc i} and CO annuli. Due to the difficulty of modeling the extra emission we reduced the number of free templates and continued the procedure with a single hydrogen template for each annulus: $N$H=$N$H~{\sc i}+2$X_{CO}$W(CO). We deduced the $X_{CO}$ conversion factor as half the ratio between the emissivity associated to the H~{\sc i} template and the one associated to the CO one. We repeated the fitting procedure with the single $N$H for inner templates  iteratively adding the positive residuals of the fit to the extra emission template. In this way we obtained a first version of the inner template emissivity and a template for the extra emission and large scale structures. \\
For the second step we fitted the whole Galaxy including the first quadrant with all the templates parameters free to vary. We added to Equation \ref{eqRing} the template for the extra emission obtained in step one. We repeated the iteration and derived the final emissivities per hydrogen atom in the inner annuli. We also obtained a template corresponding to the extra emission and large scale structures.

\begin{figure}
\begin{center}
\includegraphics[width=8cm]{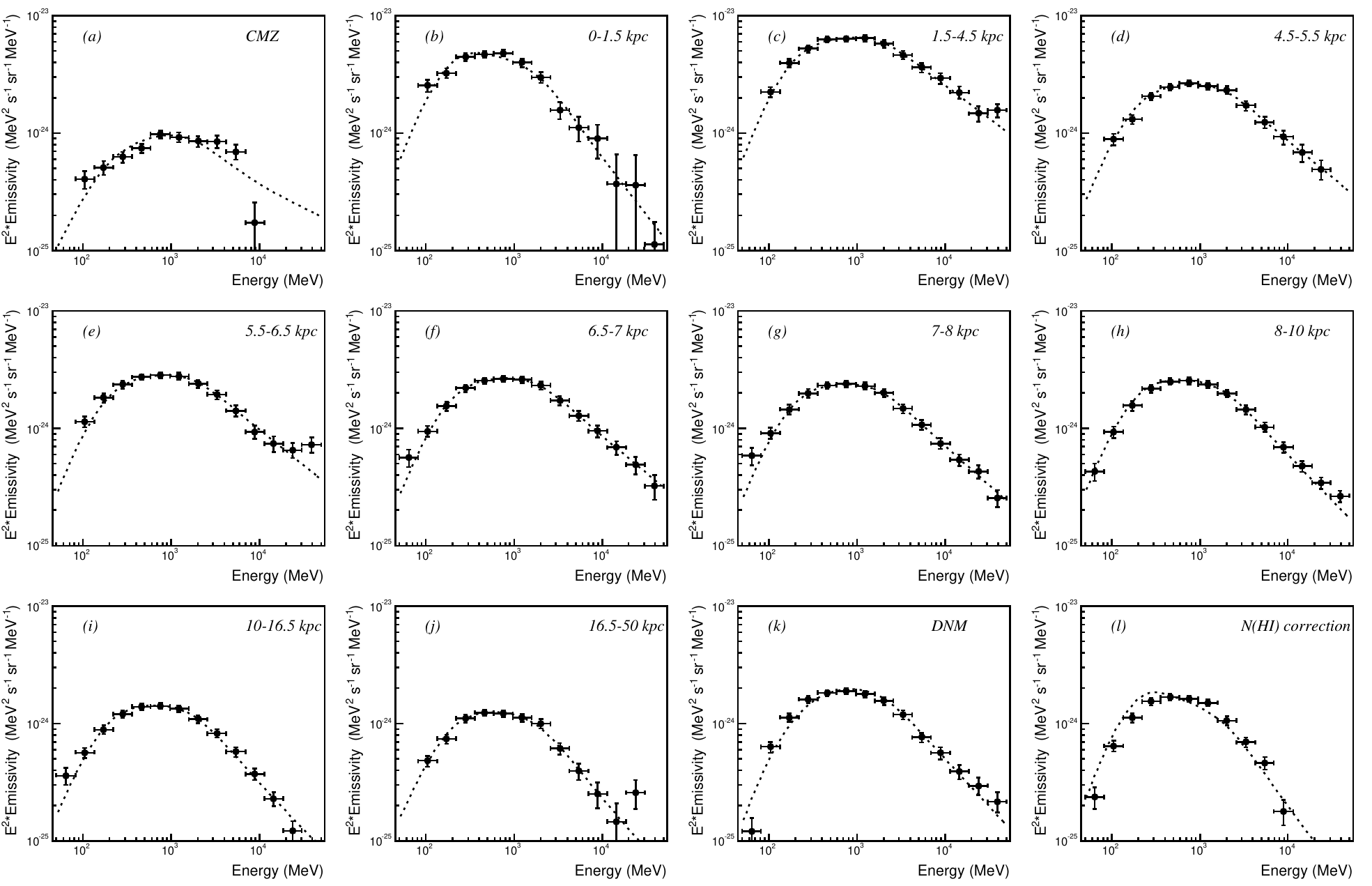}
\caption{(a)-(j): The $\gamma$-ray emissivity per hydrogen atom in H~{\sc i} and H$_2$ phases for the CMZ and the 9 Galactocentric annuli. In order to extrapolate to higher energies we fitted a model based on proton density and production cross-section (dashed line). We applied the same procedure for the DNM (k) and the $N$H~{\sc i} correction (l) maps. 
We did not display in the graph points with emissivities lower than $10^{-25} MeV^{2} s^{-1} sr^{-1} MeV^{-1}$ nor points corresponding to the lowest energy bin that were also not used in the analysis.}
\label{gas_emiss}
\end{center}
\end{figure}

Figure \ref{gas_emiss} shows the differential $\gamma$-ray emissivity per hydrogen atom $\frac{dq}{dE}=q/\Delta E$, where $\Delta E$ is the energy bin width, for the 9 Galactocentric annuli and the central molecular zone (CMZ) region scaled to emissivity per hydrogen atom assuming $X_{CO}$=0.5$\times$10$^{20}$~cm$^2$~(K~km~s$^{-1}$)$^{-1}$.
To derive an interstellar diffuse emission model at energies up to 600~GeV we fitted the emissivities with a $\gamma$-ray production model of bremsstrahlung emission and hadronic decay and used this model for the extrapolation. 
We fitted the differential emissivity of each annulus between 200~MeV and 30~GeV with a parametrized proton flux and a $\gamma$-ray production cross-section based on \cite{Kamae:2006p2590}. We adopted a spectral model for CR protons of the form: $A\beta^{P_1}R^{-P_2}$ where $\beta=v/c$, $R$ is the rigidity of the proton and ($A,P_1,P_2$) are free parameters \citep{Shikaze:2007p3492} and folded this proton functional with the $\gamma$-ray production cross-section. {\it Fermi}-LAT detects $\gamma$ rays resulting not only from proton-proton collisions but also from the interaction of heavier CR or ISM nuclei. We used the results of \cite{Mori:2009p341} to scale the proton-proton cross-section to the nucleus-nucleus cross-section taking into account the ISM and CR composition. The $\gamma$ rays detected by $Fermi$ at energies relevant for this work are also produced by bremsstrahlung of electrons and positrons. We accounted for this contribution using an electron spectral form with free parameters and the cross section of \citep{Gould:1969p2474}. 
We fitted the bremsstrahlung emission together with the hadron decay component first to the local emissivities.
We derived a proton functional parameter $P_1$ and a bremsstrahlung contribution that we assumed constant for the other annuli. We then fitted the other annuli with only two parameters: the proton spectrum normalization $A$ and proton spectral index $P_2$. In Figure \ref{gas_emiss} the emissivity resulting from this fit is represented by a dotted line.  

\begin{figure}
\begin{center}
\includegraphics[width=7cm]{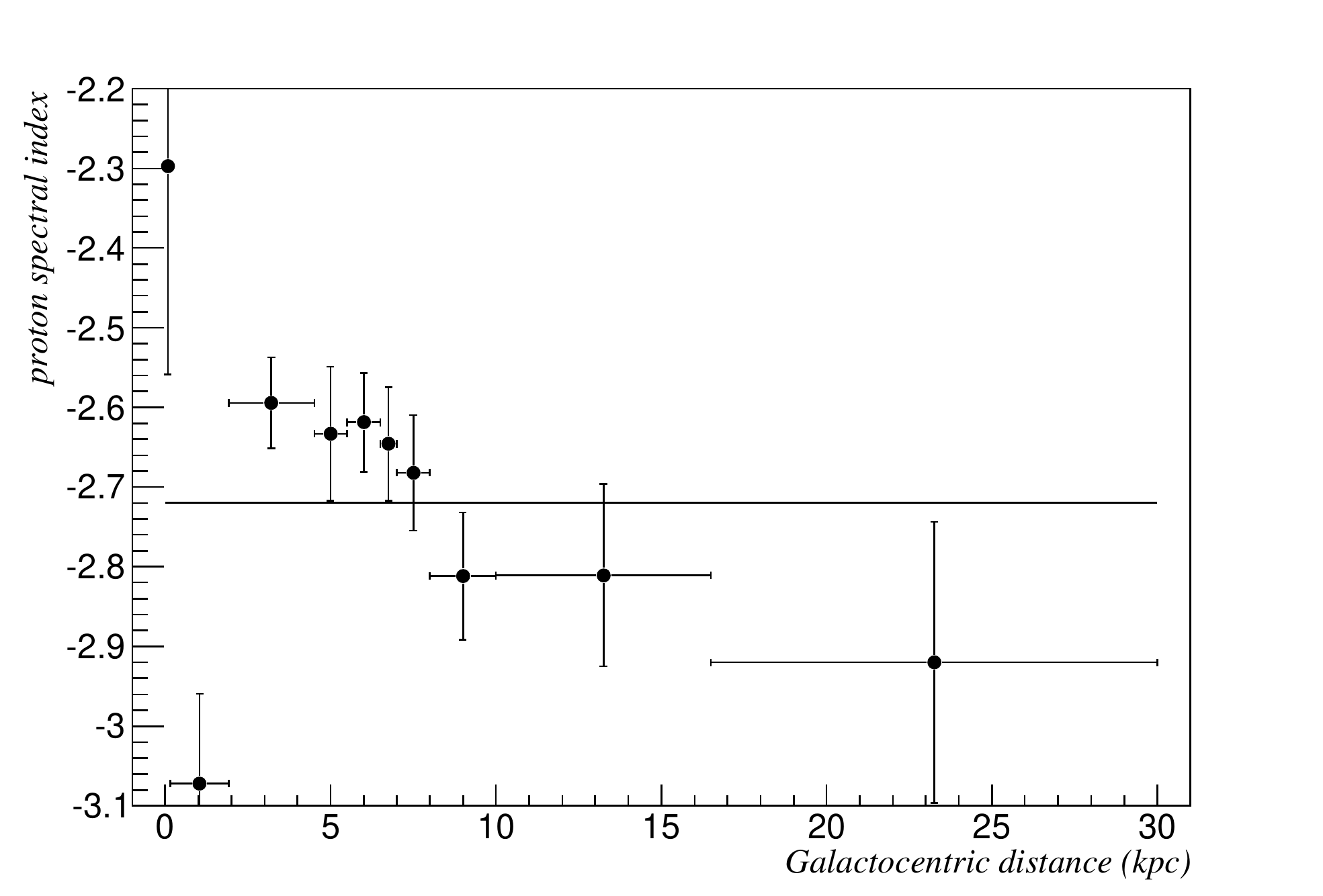}
\caption{Proton spectrum spectral index $P_2$ versus the distance from the Galaxy center. The spectral indexes are extracted from the fit of the $\gamma$-ray emissivities for various hydrogen Galactocentric annuli. We did not include any systematic uncertainty in the spectral index error bar. The Galactocentric distance error bar represents the radial width of the annuli. For comparison we draw in solid line the spectral index for proton energies above 100~GeV extracted from the GALPROP model $^SY^Z6^R30^T150^C2$ \citep{Ackermann:2012p2978}.}
\label{fig_spec_index}
\end{center}
\end{figure}

To check the validity of the $\gamma$-ray template fitting procedure we studied the coherence of the resulting proton spectral parameters between the different annuli. In Figure \ref{fig_spec_index} we plotted the proton spectral index $P_2$  versus the Galactocentric distance of the annulus. We observe a continuous softening of the proton spectra with the distance from the center of the Galaxy. The spectral indices of protons extracted from the emissivities measured in the first annulus and the CMZ correspond to a region extending $\pm$10$\degr$ from the Galactic center where confusion with other templates or point sources is possible. It might also be contaminated by the soft extra emission not totally suppressed by the iterative fit procedure. 

\begin{figure}
\begin{center}
\includegraphics[width=7cm]{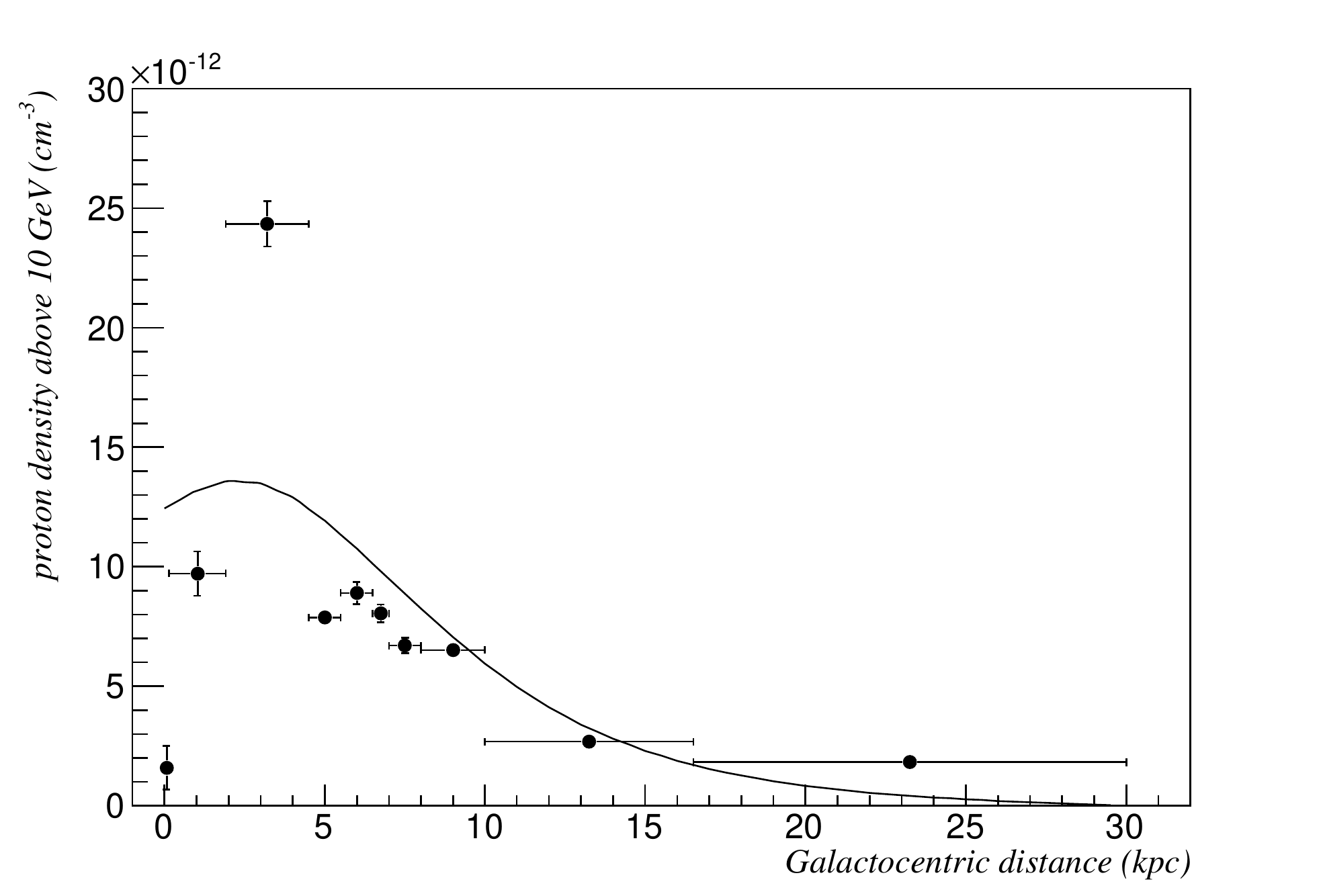}
\caption{Integrated proton flux above 10~GeV versus Galactocentric distance. For comparison we included the integrated proton flux from the GALPROP $^SY^Z6^R30^T150^C2$ model (solid line). We did not include any systematic uncertainty in the integrated proton flux error bar. Those systematic uncertainties could be significant for example close to the GC where the integrated proton flux depends on the $X_{CO}$ ratio used in the analysis.}
\label{fig_integ_flux}
\end{center}
\end{figure}

Figure \ref{fig_integ_flux} shows the radial distribution of proton density integrated above 10~GeV evaluated from the present work. 
We observe a steep CR density increase around 3~kpc. Again, part of this increase can be due to a contamination  by the extra emission of the emissivity for the annulus extending to $\pm$30$\degr$ in longitude.
We also observe that the inferred CR proton density in the CMZ is about 4 times lower than the local one (about 8 times lower if we assume the same $X_{CO}$ as for the local annuli). \cite{Blitz:1985p4121} suggested a lower $X_{CO}$ to explain the anomalously low $\gamma$-ray production compared to the CO column density in COS-B. Again, caution should be used to interpret the proton density in the CMZ given possible confusion with point sources or with the extra emission, especially at low energies. In Figure \ref{fig_integ_flux} we also show the CR proton density predicted by GALPROP $^SY^Z6^R30^T150^C2$. We note a reasonable agreement with the ones derived from $\gamma$-ray observations, however beyond 5~kpc the predicted proton density gradient is steeper than the observed one \citep{Bloemen:1986p3811,Strong:1996p1032,Abdo:2010p3307,Ackermann:2011p4125}. GALPROP also predicts a broader distribution around 3~kpc. 

We applied the same fit method for the coefficients associated with the DNM and $N$H~{\sc i} correction templates obtained from the positive and negative dust residual. In the case of the $N$H~{\sc i} correction template we left all the spectral parameters for electrons and protons free to obtain a better fit to the data. Figure \ref{gas_emiss} (k) and (l) show the emissivities associated with the DNM and the $N$H~{\sc i} correction templates together with the fit. To display in this graph the emissivities inferred from the optical depth map in the same units as those from the column density maps, we divided the measured and fitted emissivities by an arbitrary gas-to-dust ratio of $4\times10^{21}$ cm$^{-2}$ mag$^{-1}$.

From the fit of Equation \ref{eqRing} we obtained a normalization factor $N_{IC}$ continuously increasing from $\sim$1 at 50~MeV to $\sim$2 at 2~GeV and then continuously decreasing back to $\sim$1 at 50~GeV. Above 50 GeV the LAT $\gamma$-ray statistics is low causing a correlation of the IC template with the isotropic emission in the fit. We decided to rely on GALPROP predictions for the extrapolation to high energies and did not apply any scaling to $I_{IC_{p}}$ for energies above 50~GeV.

\section{Modeling the large scale structure and the extra emission}
\label{sec:modelization_of_the_large_scale_section}
\begin{figure}
\begin{center}
\includegraphics[width=8cm]{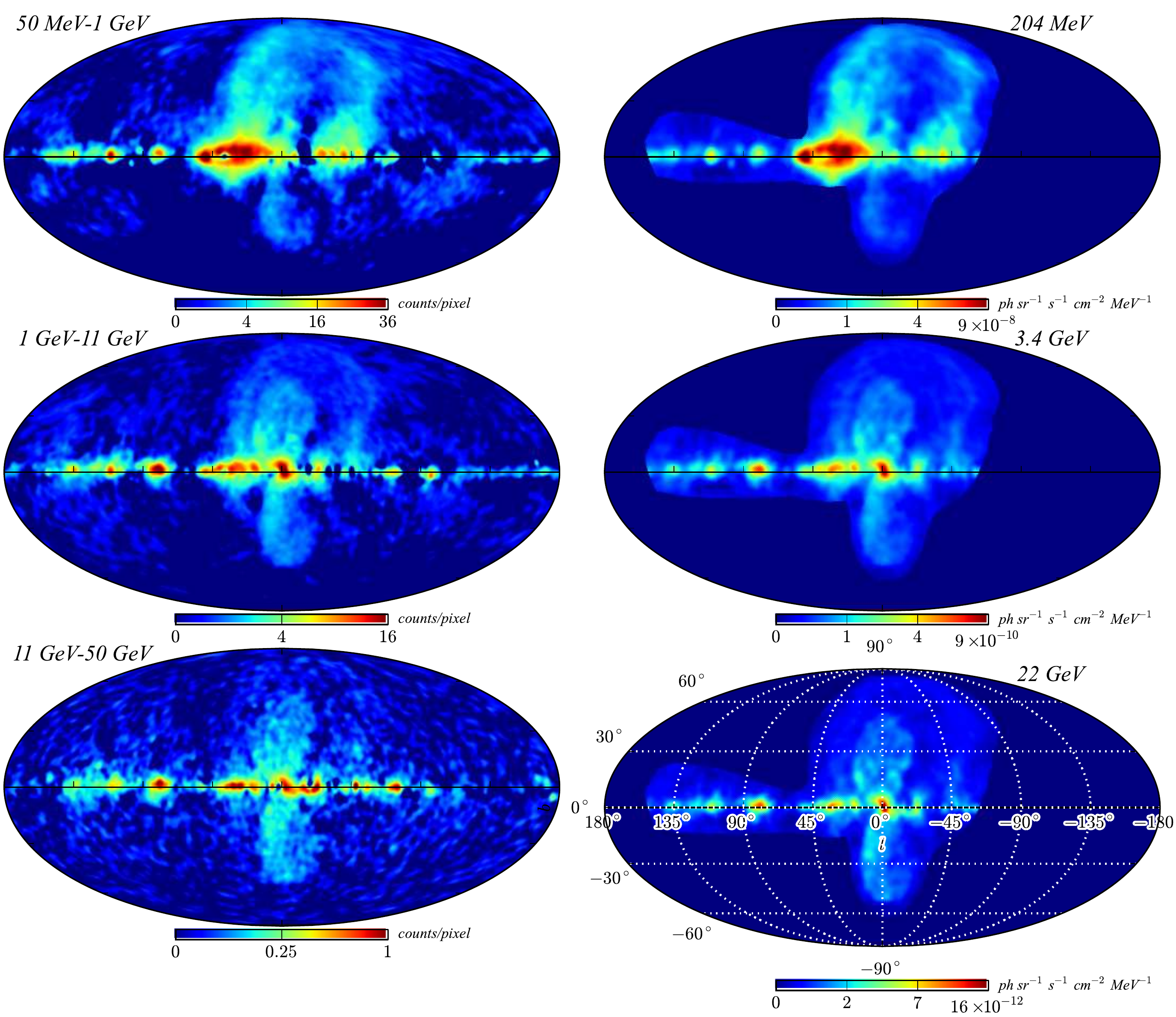}
\caption{Left column: Mollweide projection in Galactic coordinates of the {\it Fermi}-LAT counts map after subtracting point and extended sources, the limb and isotropic emission, and a conventional interstellar model based on fitted gas emissivities and scaled IC$_p$ only. The residual map is shown for three energy bands: 50~MeV--1~GeV (top), 1--11~GeV (middle), 11--50~GeV (bottom). We smoothed those three maps with a 2-dimensional symmetric Gaussian of 3$\degr$ FWHM. Right: intensity of the modeled large scale emission $RES_{IC}$ at energies: 204~MeV (top), 3.4~GeV (middle), 22~GeV (bottom). All the maps are displayed with a square root scaling and a pixel size of 0.25$\degr$.}
\label{fig_comparaison_res_model}
\end{center}
\end{figure}

\begin{figure}
\begin{center}
\includegraphics[width=7cm]{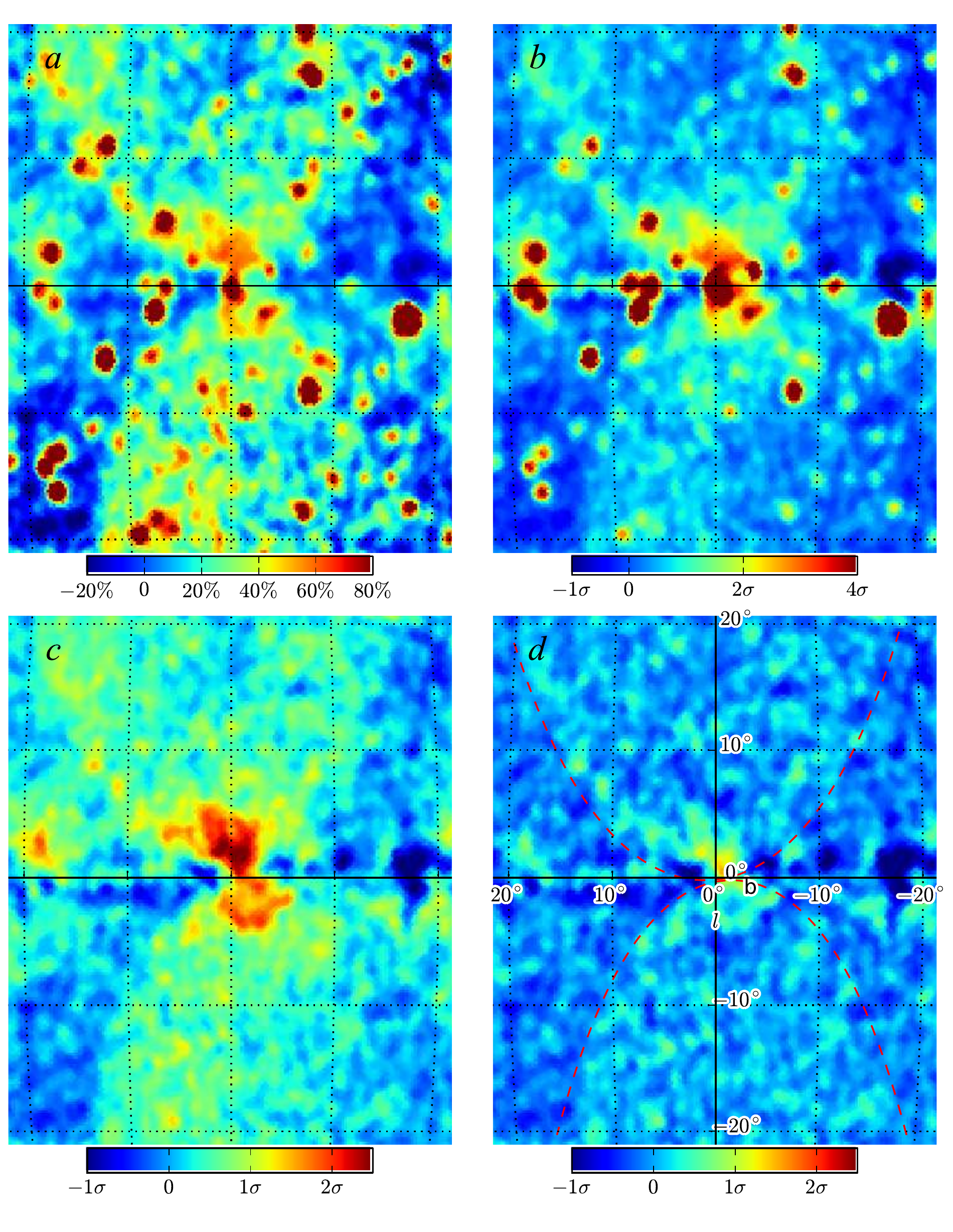}
\caption{Closeup of a region within 20$\degr$ of the Galactic center showing the {\it Fermi}-LAT counts map integrated between 1.7~{\it GeV} and 50~{\it GeV} after subtracting a conventional model of interstellar emission made only of emission correlated with the gas and the IC$_p$. To reduce the contrast between the Galactic plane emission and higher latitudes, we display the residual in fractional units, dividing the residuals by the model (a), and in units of standard deviation, dividing the residuals by the square root of the model (b). In (c) we additionally subtracted the point and extended sources from a preliminary 3FGL list. In (d) we show the residual map with the large-scale emission modeled by $RES_{IC}$ subtracted, it contains features smaller than the angular scale included in RES\_IC. The red dashed lines correspond to the catenary function $10.5\times(cosh((l-1)/10.5)-1)$ (north) and $-8.7\times(cosh((l+1.7)/8.7)-1)$ (south) that reproduce approximately the edge of the bubbles for latitude below 20$\degr$. We smoothed the 4 maps with a Gaussian of 1$\degr$ FWHM.}
\label{gal_center}
\end{center}
\end{figure}

\begin{figure}
\begin{center}
\includegraphics[width=7cm]{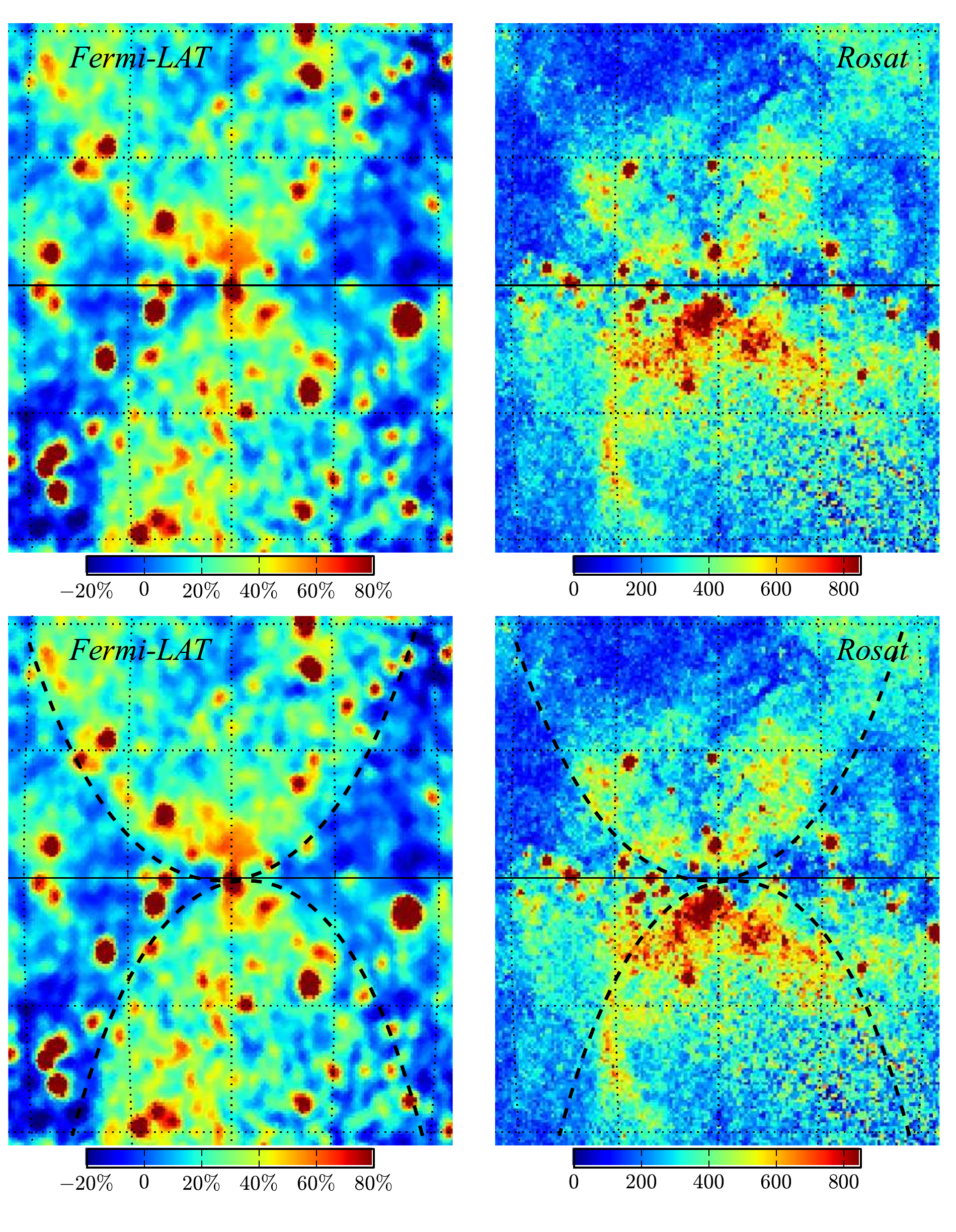}
\caption{Comparison between the {\it Fermi}-LAT residuals shown in Figure \ref{gal_center}a and the ROSAT all-sky survey \citep{Snowden:1997p3918} for energies 0.73--2.04~keV. The ROSAT observations are displayed in 10$^{-6}$~counts~s$^{-1}$~arcmin$^{-2}$. The second row shows the same plots as in the first row together with the catenary curves given in Figure \ref{gal_center}.}
\label{ROSAT}
\end{center}
\end{figure}

To include in our interstellar emission model the $\gamma$ rays produced by phenomena that lack templates like the large scale structures, we first created a conventional interstellar emission model based on gas emissivities and IC obtained as described above. We added the sources from a preliminary version of the 3FGL catalog, the predicted Sun and Moon intensities and used an isotropic and limb normalization derived from the local H~{\sc i} annuli fit. Figure \ref{fig_comparaison_res_model} (left column) shows the positive difference between the {\it Fermi}-LAT counts map and the counts map expected from this model integrated in three energy bands: 50~MeV-1~GeV, 1-11~GeV, and 11-50~GeV. As expected, since we did not include in the model any large structures nor patches, we observe positive counts residuals including the regions around Loop~{\sc i} and the {\it Fermi} Bubbles. We also observe the extra emission broadly distributed along the plane at longitudes less than 50$\degr$ and to a lesser extent at longitudes around 315$\degr$. 
We also observe an extended excess of counts toward the Galactic center at the base of the {\it Fermi} bubbles. 
Figure \ref{gal_center} shows a closeup of the Galactic center region representing the difference between the {\it Fermi}-LAT counts integrated between 1.7~GeV and 50~GeV and those expected from the conventional model using only the determined gas emissivities and IC emission. To reduce the contrast due to the bright $\gamma$-ray emission of the Galactic plane, we divided this difference by the conventional model counts (Figure \ref{gal_center}a) and by the square root of this number (Figure \ref{gal_center}b,c,d). The fluxes of some preliminary 3FGL sources located in the Galactic ridge depend to some extent on the interstellar emission model, so a fraction of the interstellar emission can be incorrectly assigned to point sources. To avoid any bias  we did not subtract them from the counts map in Figure \ref{gal_center}a and \ref{gal_center}b.  They were subtracted in Figure \ref{gal_center}c and \ref{gal_center}d. 
We deduce from those plots that the bases of the {\it Fermi} bubbles have the form of a catenary in which the $\gamma$-ray emission is enhanced. We observe hints that this enhancement is perpendicular to the Galactic plane and originates from the $\gamma$-ray source located in the direction of the Galactic center. 
As pointed out by \cite{Su:2010p2675} the ROSAT all-sky survey \citep{Snowden:1997p3918} shows structures similar to the one of the {\it Fermi} bubbles. Figure \ref{ROSAT} shows that the {\it Fermi} bubbles have a similar shape within 20$\degr$ from the Galactic center in X and $\gamma$ rays. We note that the X-ray detected by ROSAT are strongly absorbed at absolute latitudes of less that 2$\degr$ which can produce the same artifacts as the use of an overestimated hydrogen emissivities for the inner annuli in the {\it Fermi}-LAT residual maps.
Other extended excesses are present along the Galactic plane including in the Cygnus region. We did not observe strong negative residuals except in the direction of the Carina arm tangent where the model largely over-predicts the observations.

We chose to model those large scale and extra $\gamma$-ray emissions by assuming they all originate from IC interactions of a population of CR electrons with the cosmic microwave background radiation. The physical motivation behind the production of those $\gamma$ rays in the energy range between 50~MeV and 50~GeV is not relevant if the resulting model is consistent with the data. For simplicity we favored a unique ISRF and $\gamma$-ray production process. We fitted in each pixel an electron spectral form so that the conventional model added to this IC emission reproduces the total counts of the pixel. In order to reduce in the modeled IC map the number of undetected point sources and small extended structures coming from the counts map, we transformed the spatial distribution of the electron spectral parameters into wavelets and filtered out scales smaller than 2$\degr$. We created a spatial mask drawn by hand that encompasses only regions where the large-scale structures and extra emission is the largest. We call this masked and filtered emission $RES_{IC}$ and show it in the right column of Figure \ref{fig_comparaison_res_model} at energies approximately corresponding to the geometric average of the energy interval used to display the counts map of the left column. We observe a good agreement between the large-scale structures observed in the counts map and the one we have modeled. In Figure \ref{gal_center}d we show the residual map integrated between 1.7~GeV and 50~GeV in the direction of the Galactic center when $RES_{IC}$ is included in the model. The residual map is flat apart from some emission toward the Galactic center and the {\it Fermi} bubbles. This emission corresponds to the small scales filtered out in the wavelet decomposition.

\section{Resulting model of Galactic interstellar  emission}
\label{sec:Resulting_model_section}
We derived a final model for the interstellar emission from the sum of the modeled differential gas $\gamma$-ray emissivities ($\frac{dq_{fit}}{dE}$), the renormalized $I_{IC_{p}}$, and from the large-scale and extra emission $RES_{IC}$ (Equation \ref{eqring3}).  

{\footnotesize
\begin{equation}
\begin{split}
  I(E,l,b) =  \sum_{i=HI,H_{2},DNM} \frac{dq_{fit_{i}}}{dE}(E) I_{H_i}(l,b) \\  +  {N}_{IC}(E)I_{IC_{p}}(E,l,b) + RES_{IC}(E,l,b)
\end{split}
\label{eqring3}
\end{equation}
}

\begin{figure}
\begin{center}
\includegraphics[width=7cm]{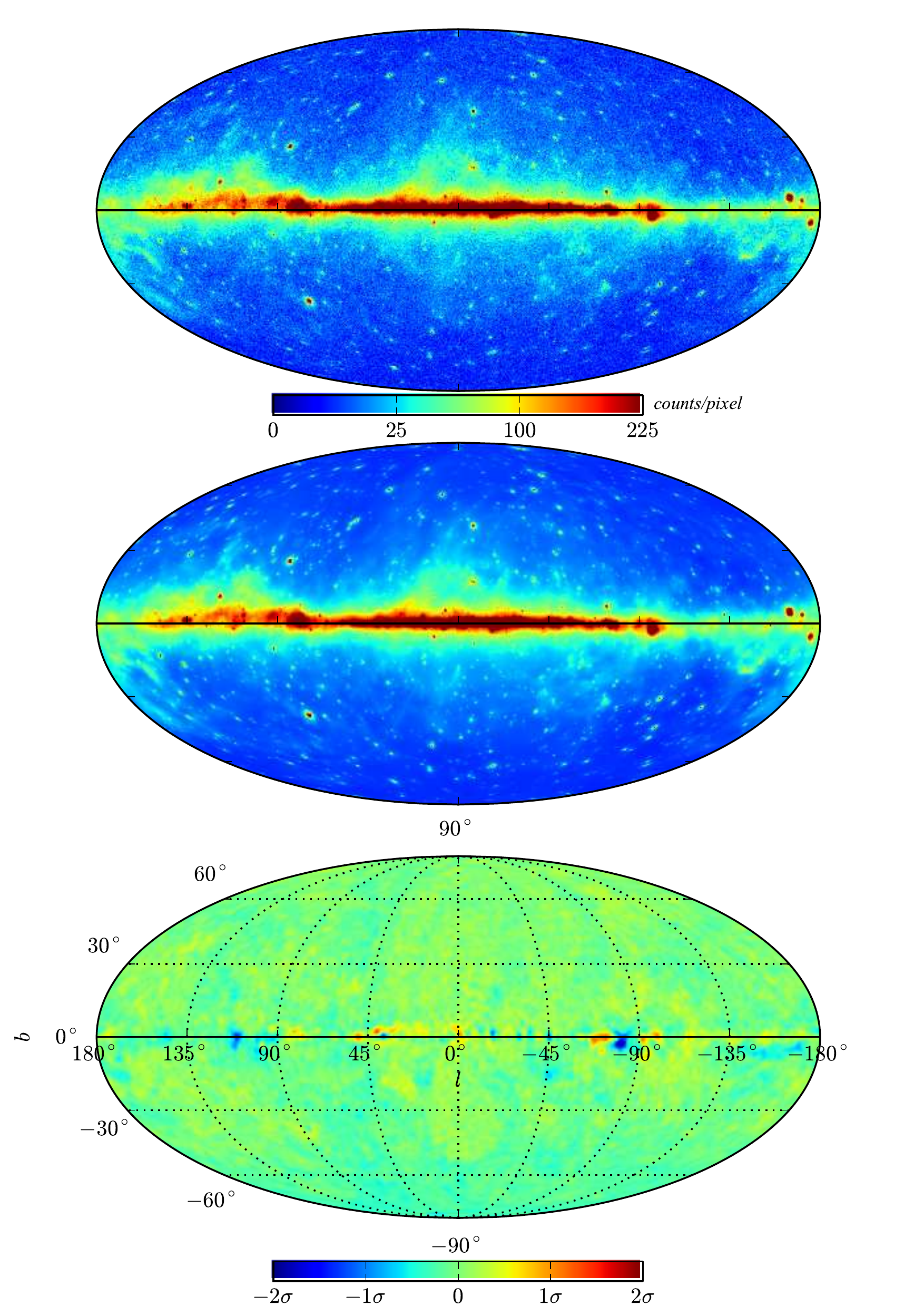}
\caption{Top: All-sky Mollweide projection for 4 years of {\it Fermi}-LAT $\gamma$-ray counts in the 0.36--50~GeV energy band. Middle: counts prediction in the same energy range based on the interstellar emission model combined with modeled point and extended sources (including the Sun and the Moon), the Earth limb emission and the isotropic emission. Both maps are displayed with square root scaling to enhance emission away from the plane. Bottom: residual map in units of standard deviations after smoothing with a Gaussian of 2$\degr$ FWHM. The pixel size for the three maps is 0.25$\degr$.}
\label{counts}
\end{center}
\end{figure}

We compared the {\it Fermi}-LAT counts map integrated between 360~MeV and 50~GeV (Figure \ref{counts}, top) to the one predicted by the interstellar emission model given by Equation \ref{eqring3} combined with the ones originating from non-Galactic interstellar origin (Figure \ref{counts}, middle) . We derived the residual map (Figure \ref{counts}, bottom) by subtracting the model from the data and normalizing by the square root of the model to enhance deviations above statistical fluctuations. The overall agreement between observations and model is very good, partly because some excesses we observed were modeled and re-injected into the interstellar model.

We derived from Equation \ref{eqring3} a model of the interstellar emission available at the {\it Fermi} Science Support Center (FSSC) website as a FITS file named {\it gll\_iem\_v06.fit}. We resampled all the maps to a 0.125$\degr$ grid. The FITS file comprises 30 logarithmically-spaced energy bins between 50 MeV and 600 GeV. It gives the differential intensity of the Galactic diffuse emission model in photons sr$^{-1}$ s$^{-1}$ cm$^{-2}$ MeV$^{-1}$. This model tuned to LAT data is not corrected for the energy dispersion, it can then be used directly with LAT data. The Pass 7\_V15 IRFs (P7REP\_CLEAN\_V15) are the recommended set for Pass 7 reprocessed data and this model. The difference with the Pass 7\_V10 used for this fitting mainly resides in an improved Monte Carlo PSF and in an updated fitting procedures to determine the parameters for the LAT effective area representation. Those minor differences modify the exposure but not the reconstructed LAT events. The minimum ratio of exposure maps (V15/V10) is 0.98 at 50~MeV and the maximum 1.05 at 1~GeV. In order to use the model with the final IRFs we rescaled the intensity by the ratio of the exposure maps evaluated for each of the 30 energy bins of the Galactic diffuse emission model. The model is then intended for use with the instrument response functions versions P7REP\_SOURCE\_V15, P7REP\_CLEAN\_V15, and P7REP\_ULTRACLEAN\_V15.

\acknowledgments
The \textit{Fermi} LAT Collaboration acknowledges generous ongoing support
from a number of agencies and institutes that have supported both the
development and the operation of the LAT as well as scientific data analysis.
These include the National Aeronautics and Space Administration and the
Department of Energy in the United States, the Commissariat \`a l'Energie Atomique
and the Centre National de la Recherche Scientifique / Institut National de Physique
Nucl\'eaire et de Physique des Particules in France, the Agenzia Spaziale Italiana
and the Istituto Nazionale di Fisica Nucleare in Italy, the Ministry of Education,
Culture, Sports, Science and Technology (MEXT), High Energy Accelerator Research
Organization (KEK) and Japan Aerospace Exploration Agency (JAXA) in Japan, and
the K.~A.~Wallenberg Foundation, the Swedish Research Council and the
Swedish National Space Board in Sweden.
 
Additional support for science analysis during the operations phase is gratefully acknowledged from the Istituto Nazionale di Astrofisica in Italy and the Centre National d'\'Etudes Spatiales in France.

Some of the results in this paper have been derived using the HEALPix \citep{Gorski:2005p1076} package.

\bibliography{my_paper_template}

\end{document}